\documentclass[preprint,10pt]{elsarticle}
\usepackage{amsfonts,amsmath,amssymb,amsthm}
\usepackage{graphicx}
\usepackage{subfigure}
\begin{document}

\begin{frontmatter}

\title{A Method of Measuring Cosmic Magnetic Fields with Ultra High Energy Cosmic Ray Data}

\author{Martin Erdmann}
\author{Peter Schiffer}
\address{Physics Institute 3A, RWTH Aachen University, 52056 Aachen, Germany}
\ead{schiffer@physik.rwth-aachen.de}


\begin{abstract}
We present a method to measure cosmic magnetic fields with ultra high energy cosmic rays (UHECRs). We apply an advanced autocorrelation method to simulated UHECRs which includes their directional as well as energy information. Without explicit knowledge of the UHECR sources, such measurements are sensitive to the number of sources and to the magnetic field strength subjected to the UHECRs. Using a UHECR Monte Carlo model including sources, random walk propagation and a coherent deflection, we explain the procedure of reconstructing the allowed phase space of the model parameters from a simulated autocorrelation measurement.

\end{abstract}

\end{frontmatter}

\section{Introduction}

Recent results of the Pierre Auger Observatory \cite{AGN, AGNlong} show a correlation of ultra high energy cosmic rays (UHECRs) above 56~EeV with nearby active galactic nuclei (AGN) at distances up to 75~Mpc. The exact origin of the UHECRs and the positions of their sources still remain unknown but it is favored that UHECRs are accelerated at discrete sources.

The UHECR’s propagate through extragalactic and galactic magnetic fields, and consequently carry information about the direction and strength of the fields. For UHECR’s originating from a single source with energies above a certain threshold, the influence of the fields is expected to result in an energy ordering with respect to the direction of the source.

Galactic magnetic fields are known to some extent, from different measurement methods. For recent reviews refer to, e.g. \cite{Beck, Han}. Information about extragalactic fields exists for a number of galaxy clusters, however, the knowledge of magnetic fields in filaments is uncertain to at least three orders of magnitude \cite{Sigl, Dolag, Ryu}. Measurements of UHECR’s therefore have good potential of providing additional information on these magnetic fields.

In this contribution we present a new method of obtaining information on the magnetic fields without prior knowledge of the source positions. This is achieved by a statistical approach based on the concept of energy dependent angular ordering of the UHECRs mentioned above. This publication is organized as follows. First we define an energy-energy-correlation observable used to measure the angular ordering. Then we explain the analysis procedure, and demonstrate its application using a Monte Carlo generated UHECR scenario. Finally, we present a method to test models of UHECR emission and propagation using a simulated energy-energy-correlation measurement. In this context we reconstruct the parameters of a simple magnetic field model which are the number of sources emitting UHECR’s, and the turbulent magnetic field strength, while keeping a coherent magnetic field strength constant.

\section{Definition of Energy-Energy-Correlations}\label{sec:EEC}
Energy-energy-correlations are a well known quantity from high energy phys\-ics, see e.g. \cite{H1}. Their definition usually includes the energies and the angular distances of particles, which enables investigation of energy ordering in the UHECR sky. We define the energy-energy-correlation $\Omega_{ij}$ between the UHECRs $i$ and $j$ by

\begin{equation}\label{eq:omega}
 \Omega_{ij}=\frac{\left(E_i(\alpha_i)-\left<E(\alpha_i)\right> \right)  \left(E_j(\alpha_j) -\left<E(\alpha_j)\right> \right) }{ E_i(\alpha_i) E_j(\alpha_j)}.
\end{equation}

\begin{figure}
 \centering
 \includegraphics[scale=0.7]{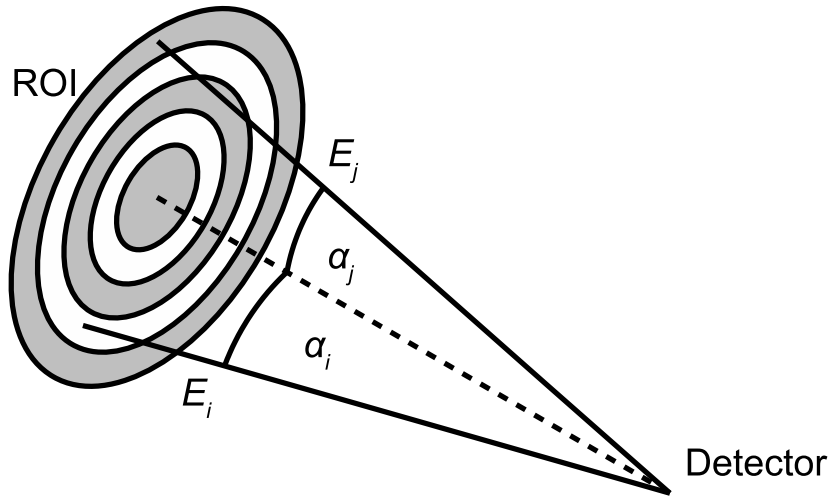}
 \caption{Schematic view of energy-energy-correlations}
 \label{fig:f1}
\end{figure}

Here $E_i$ is the energy of the UHECR $i$, and $\alpha_i$ denotes its angular distance with respect to the center of a region of interest (ROI) (figure \ref{fig:f1}). The ROI covers a limited solid angle and will be precisely outlined in section~\ref{sec:roi}. $\langle E(\alpha_i) \rangle$ denotes the corresponding average energy value of the UHECRs arriving within the same ring interval relative to the ROI center.  

Using $\Omega_{ij}$, angular ordering of the UHECRs is measured in the following sense,

\begin{enumerate}

\item

A pair of UHECR’s, one being above and the other below the corresponding average energy values, results in a negative correlation $\Omega_{ij} < 0$. This is a typical case for a background contribution.

\item

A pair with both UHECRs having energy values above or below the average energy value at their corresponding angular distance gives a positive correlation $\Omega_{ij} > 0$. Here signal UHECRs are expected to contribute.

\end{enumerate}

This means we expect overall larger $\Omega_{ij}$ for a mixing of coherently arriving UHECRs from a source with background UHECRs, than for an exclusively isotropic arrival.
In this way, the energy-energy-correlations as defined above quantify the coherence in angular ordering of the UHECRs coming from the same source, and provide a separation of incoherently arriving UHECRs from different sources on a statistical basis.

\section{Analysis Method}\label{sec:analysis}

\subsection{Signal Data Set}

In order to evaluate the sensitivity of the energy-energy-correlation observable, we use instead of real data, a set of simulated UHECRs resulting from a source distribution. We will reference this in the following as signal data set.
 
Future Monte Carlo simulations of individual UHECRs aim to include specific candidate sources, a realistic distribution of galaxies, galaxy clusters and filaments, and a proper propagation through the corresponding media and magnetic fields. Here we use a simple representation of such a UHECR Monte Carlo generator which includes a random walk propagation and a coherent deflection of the UHECRs. Random walk propagation can be considered to approximate the influence of extragalactic magnetic fields (EGMF) \cite{Waxman}, while it is assumed that the dominating effect of the galactic magnetic field (GMF) is a coherent deflection \cite{Stanev}. 

The random walk propagation is simulated by varying the angle $\theta$ between the source direction and the arrival direction of the UHECR. $\theta$ depends on the distance $D$ which a proton propagates through magnetic fields of strength $B$ in zones of the correlation length $\lambda$. The angle is chosen randomly according to a 2-dimensional Gaussian distribution with the width

\begin{equation}\label{eq:prop} 
 \sigma_\theta \simeq 0.025\deg  \left(\frac{D}{\lambda}\right)^{1/2} \left(\frac{\lambda}{10 \mathrm{\ Mpc}}\right) \left(\frac{B}{10^{-11} \mathrm{\ G}}\right)\left(\frac{E}{10^{20}\mathrm{\ eV}}\right)^{-1} .
\end{equation}

For the GMF we adopt a homogeneous magnetic field perpendicular to the galactic plane as a simple representative model which simulates its most important features. In galactic coordinates this results in the following deflection,


\begin{equation}
 \Delta \ell \simeq C_{Coherent Field} \left( \frac{10^{18} \mathrm{eV}}{E} \right).
\end{equation}

Every UHECR is shifted in galactic longitude $\ell$ depending on its energy by the amount $\Delta \ell$. The galactic latitude remains unchanged. Following from this the coherent deflection strength is maximal near the galactic plane, while it drops to zero at the poles. Although this gives a completely different global deflection pattern of UHECRs compared to e.g. \cite{Stanev}, the local effect of an energy ordering is the same. For the signal data set as well as for all other Monte Carlo sets used in this paper we adopt a coherent deflection strength of $C_{Coherent Field}=10~\mathrm{rad}$. In case of an UHECR of $60 \mathrm{EeV}$ this corresponds to a maximal deflection of $~10^\circ$, and an average deflection of $~7.5^\circ$ which is compatible with the results of \cite{Stanev}.

For the energy distribution of the UHECRs we follow the spectrum measured by the Pierre Auger Observatory \cite{spec},

\begin{equation}
 J(E)\propto E^{\gamma}
\end{equation}
with $\gamma=-2.69$ for $4\mathrm{\ EeV}<E<40\mathrm{\ EeV}$, $\gamma=-4.2$ for $E>40\mathrm{\ EeV}$.

In the production of our signal data set we first generated 10 randomly distributed sources in the sky. We assumed that all sources have a similar distance to the earth such that equation \ref{eq:prop} simplifies to

\begin{equation}\label{eq:prop2}
 \sigma_\theta \simeq C_{Random Field}\left(\frac{E}{10^{18}\mathrm{eV}}\right)^{-1} .
\end{equation}

\begin{figure}
\label{fig:f2}
 \centering
 \subfigure[]{
 \includegraphics[scale=0.27]{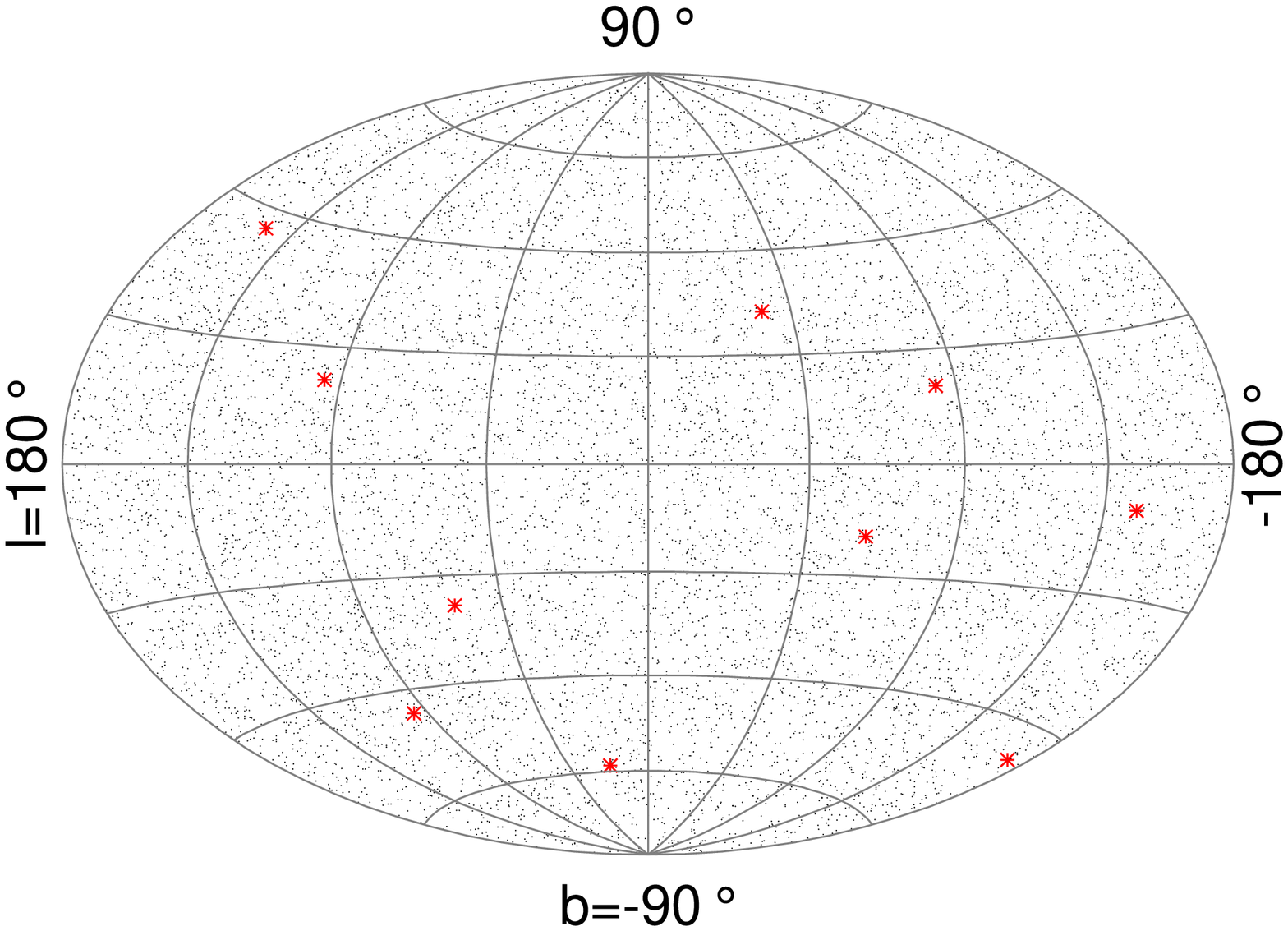}
 \label{fig:f2a}
 }
 \qquad
 \subfigure[]{
 \includegraphics[scale=0.27]{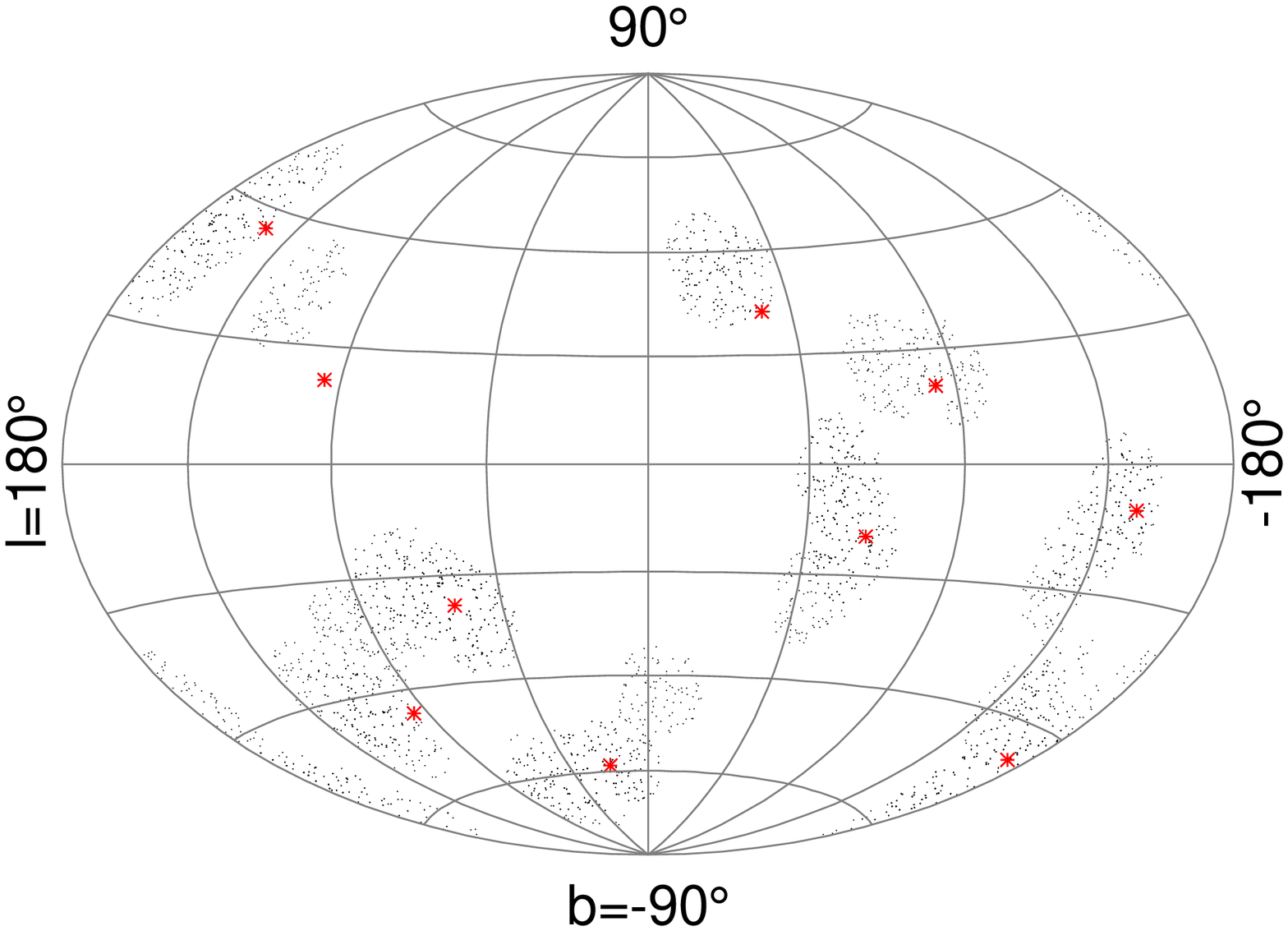}
 \label{fig:f2b}
 }\\
 \subfigure[]{
 \includegraphics[scale=0.27]{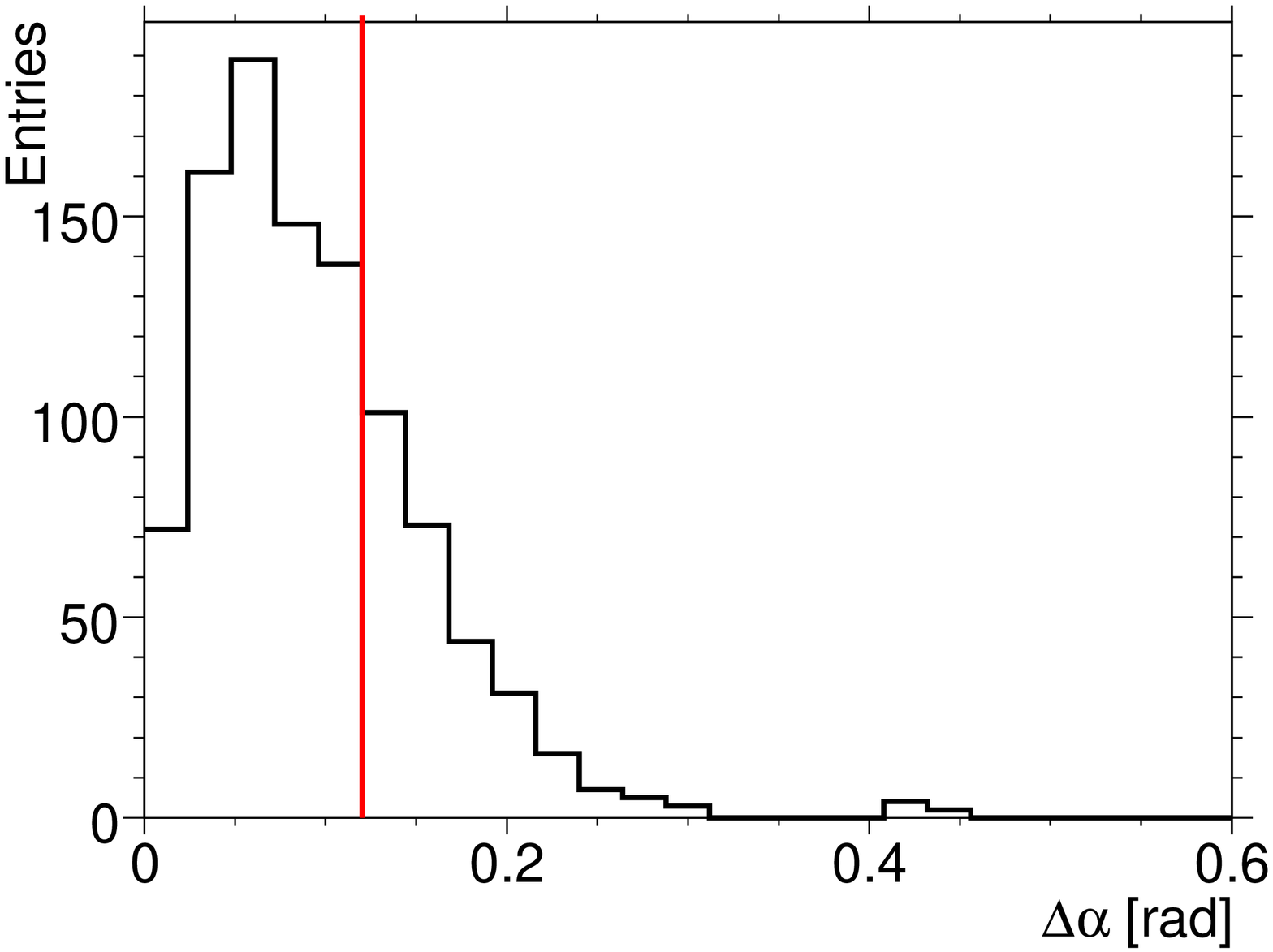}
 \label{fig:f2c}
 }
 \qquad
 \subfigure[]{
 \includegraphics[scale=0.27]{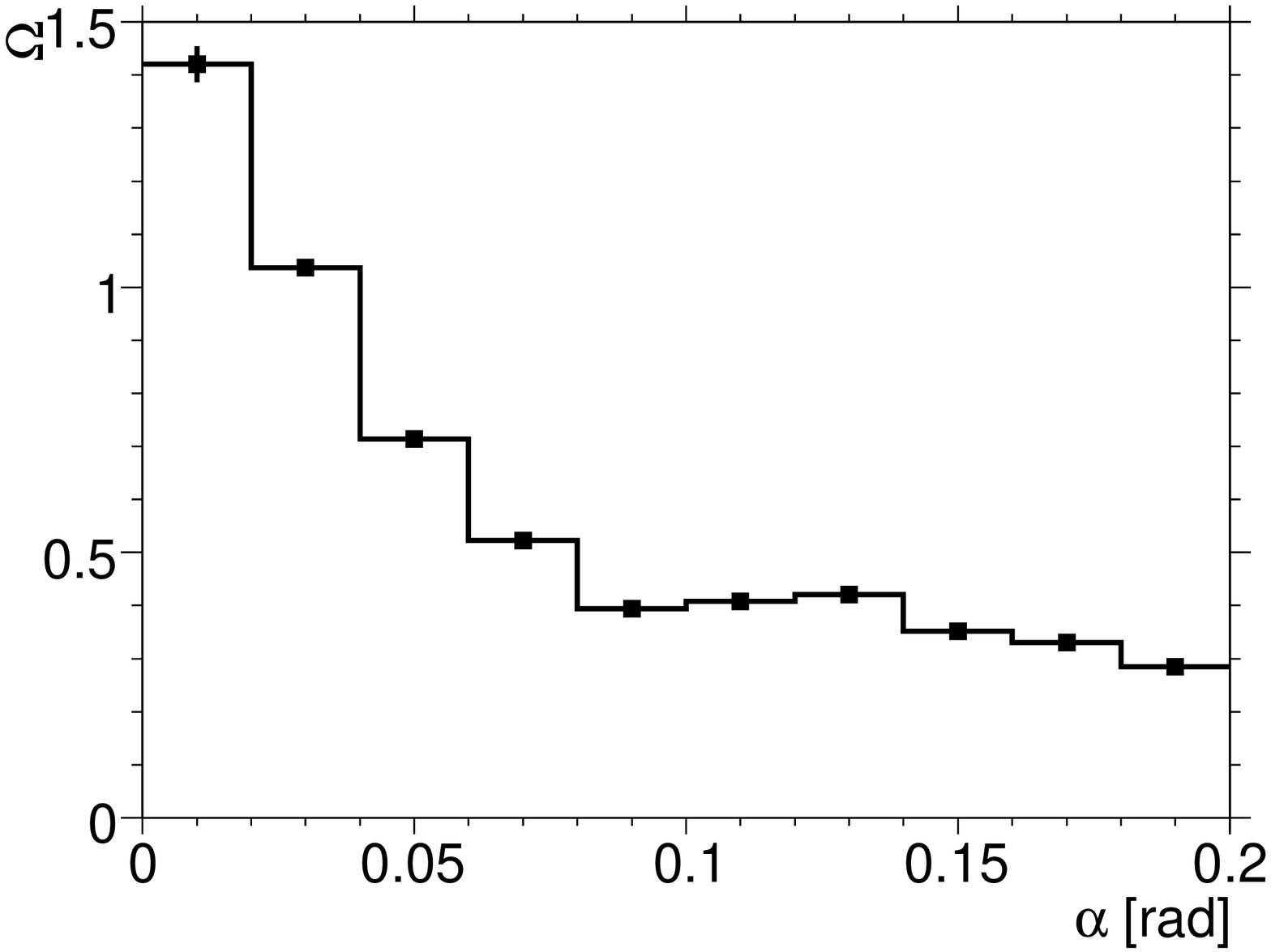}
 \label{fig:f2d}
 }
 
 \caption{\subref{fig:f2a} Arrival directions of the signal data set (black point symbols) and the sources (red asterisk symbols), \subref{fig:f2b} UHECRs belonging to regions of interest (black point symbols) and the sources (red asterisk symbols), \subref{fig:f2c} Reconstruction quality of the source direction using the cone algorithm (red line: 68\%-quantile), \subref{fig:f2d} Energy-energy-correlations of the signal data set}
\end{figure}

We produced the signal data set with $C_{Random Field}=10$~rad. A scenario compatible with this choice is, e.g. $D=50$~Mpc, $\lambda=1$~Mpc, and $B=3$~nG. We further assumed that all sources have identical luminosity, and each of them emitted 1000 protons. The total signal data set amounts to 10000 UHECRs with energies above 5~EeV. This energy threshold has been chosen to be above the ankle measured in \cite{spec}, such that the UHECRs can be considered to have extragalactic origin \cite{ankle}. In figure \ref{fig:f2a} we show the positions of the sources (asterisk symbols), and the resulting UHECR distribution (point symbols) as observed at the earth using the Hammer projection of the galactic coordinates. The UHECR distribution appears to be almost isotropic.

\subsection{Regions of Interest} \label{sec:roi}

We define a ROI as a region that is close to a source candidate of UHECRs. Since no UHECR source has been clearly identified so far, we applied a simple iterative cone algorithm to find ROI's that have an increased probability to contain a source. The algorithm works as follows.

\begin{enumerate}
\item Select all UHECRs with energies above $E_{min}=60$~EeV as initial seeds,
\item For all seeds, define a corresponding ROI by assigning all UHECRs with angular distances less than $\alpha_{max}=0.2$~rad,
\item Calculate the center of mass of each ROI using the energies of the UHECRs as weights,
\item Use the center of mass of the ROI as a new seed, and iterate starting from item 2.
\end{enumerate}
Note that in this algorithm every UHECR can be part of several ROIs. The algorithm is processed in total three times, and the last resulting ROIs are taken for further analysis.

In order to test the accuracy of reconstruction of a source direction with the ROI method, we produced 100 additional simulated data sets using the same magnetic field strength, and the same number of sources as above. We varied the random generator seeds to produce different source positions and UHECRs.

The performance of the ROI algorithm is shown in figures \ref{fig:f2b} and \ref{fig:f2c}. In figure~\ref{fig:f2b} one can see, that in the signal data set 9 out of 10 sources are covered by ROIs. Using the 100 additional data sets with the identical parameters, 95\% of the sources are covered by ROIs. In figure~\ref{fig:f2c}, the angular distance $\Delta \alpha$ between a source and the closest ROI is shown. We obtain an angular resolution in terms of the 68\%-quantile of 0.12~rad. Note that some reconstructed ROIs do not correspond to sources, and provide a background contribution to the following analysis.

\subsection{Calculation of Energy-Energy-Correlations}

In the next step we calculate the energy-energy-correlations of all pairs of UHECRs that belong to the same ROI. Each value of $\Omega_{ij}$ (equation \ref{eq:omega}) is filled into a histogram at both angular distances $\alpha_i$ and  $\alpha_j$. Finally, in every angular bin of $\alpha$ we calculate the average value $\Omega$ and its uncertainty. In figure \ref{fig:f2d}, we show the distribution of $\Omega$ where we have included the $\Omega_{ij}$ values of all ROIs simultaneously.
The corresponding error bars are small compared to the size of the symbols.
The result is the angular distribution of the energy-energy-correlations, which represents an estimator of coherence of the UHECR sky.

In experiments measuring UHECRs, this distribution can be obtained directly from the data
without any assumption on sources, propagation, or magnetic fields.

\section{Evaluation of UHECR models}\label{sec:reco}

Each state-of-the-art model of UHECR emission and propagation can be  
confronted with the distribution of an energy-energy-correlation measurement as presented above. 
In this section we present a method of evaluating such a model using the data distribution.
To explain the evaluation procedure we use simplified models of the UHECR sky.

As an example, we demonstrate the potential to exclude a model of isotropic arrival directions by comparison with the signal data distribution shown in figure~\ref{fig:f2d}. Furthermore, we evaluate variants of the above mentioned magnetic field model using different numbers of sources and magnetic field parameters. We demonstrate how phase space regions of possible parameter settings of this model can be excluded. Finally we quantify the precision of the reconstruction of these parameters.

\begin{figure}
\label{fig:f3}
 \centering
 \subfigure[]{
 \includegraphics[scale=0.27]{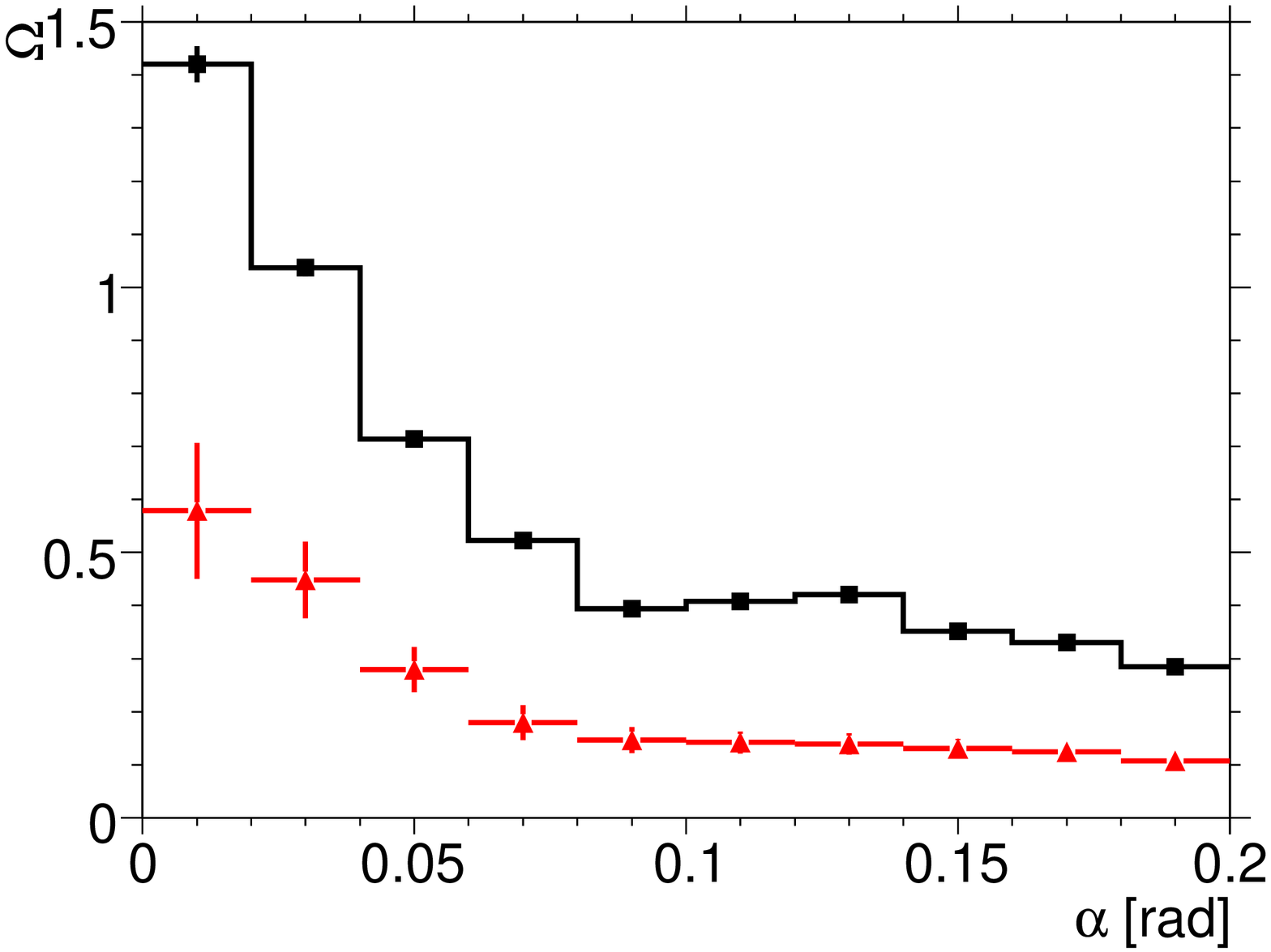}
 \label{fig:f3a}
 }
 \qquad
 \subfigure[]{
 \includegraphics[scale=0.27]{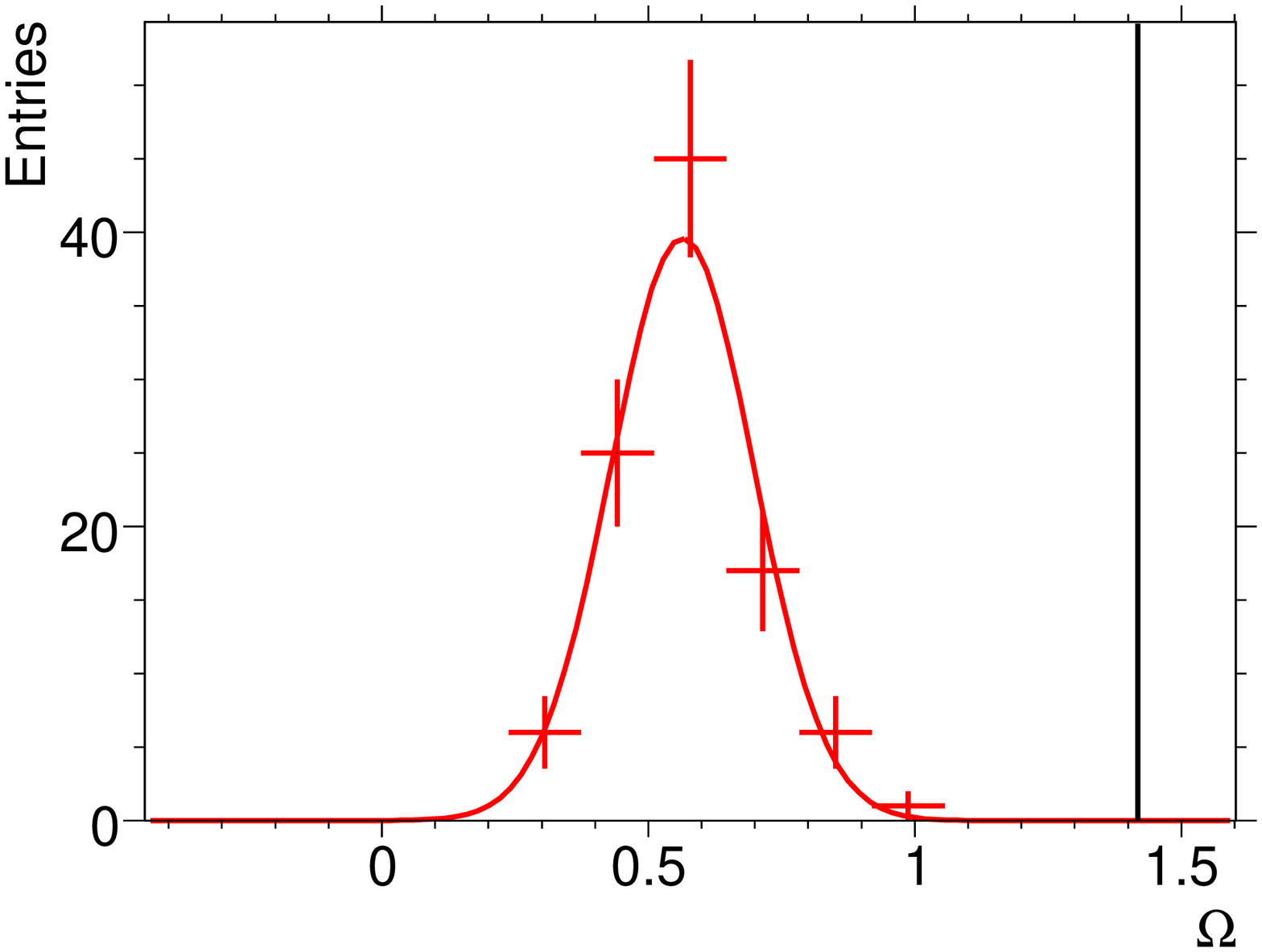}
 \label{fig:f3b}
 }\\
 \subfigure[]{
 \includegraphics[scale=0.27]{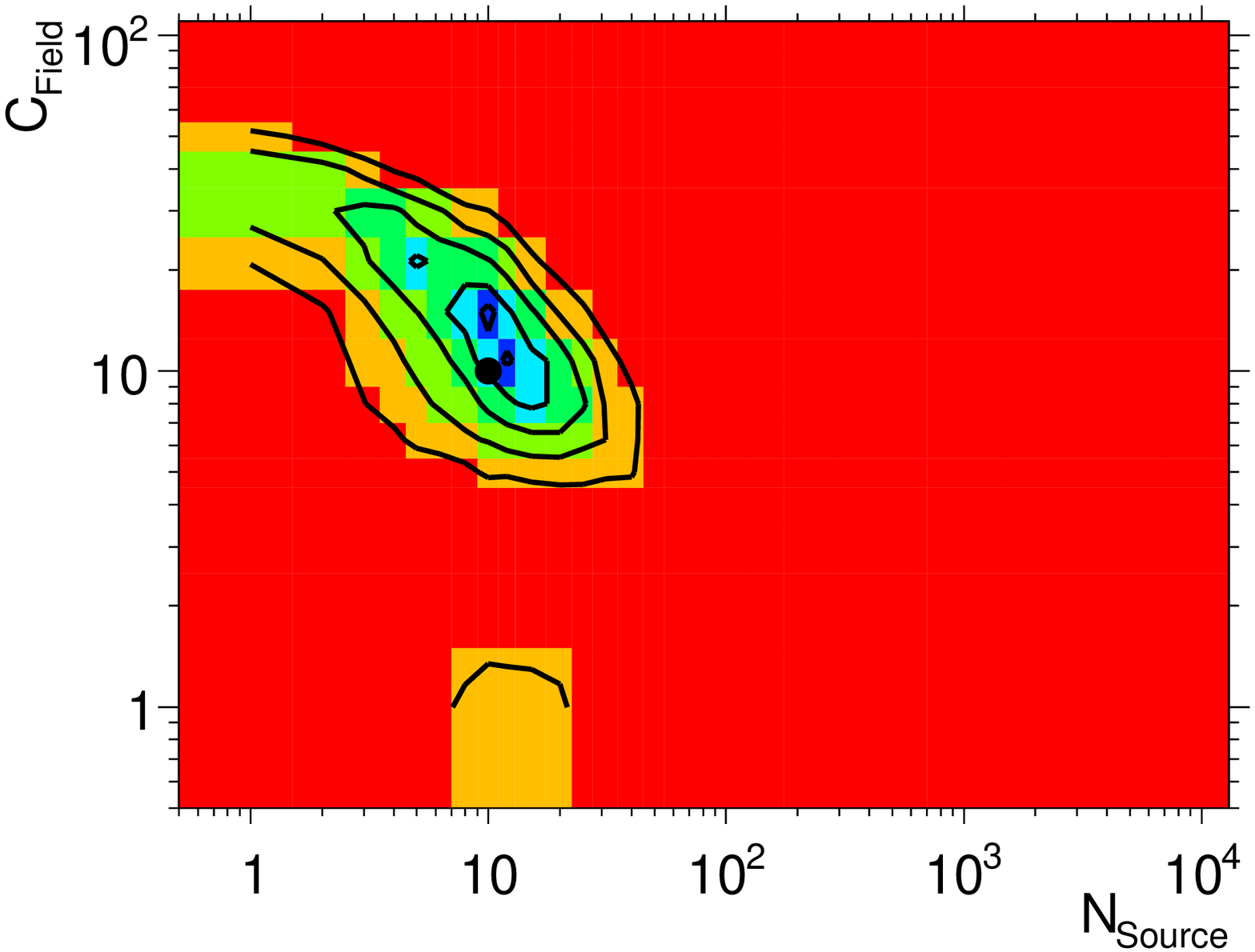}
 \label{fig:f3c}
 }
 \qquad
 \subfigure[]{
 \includegraphics[scale=0.27]{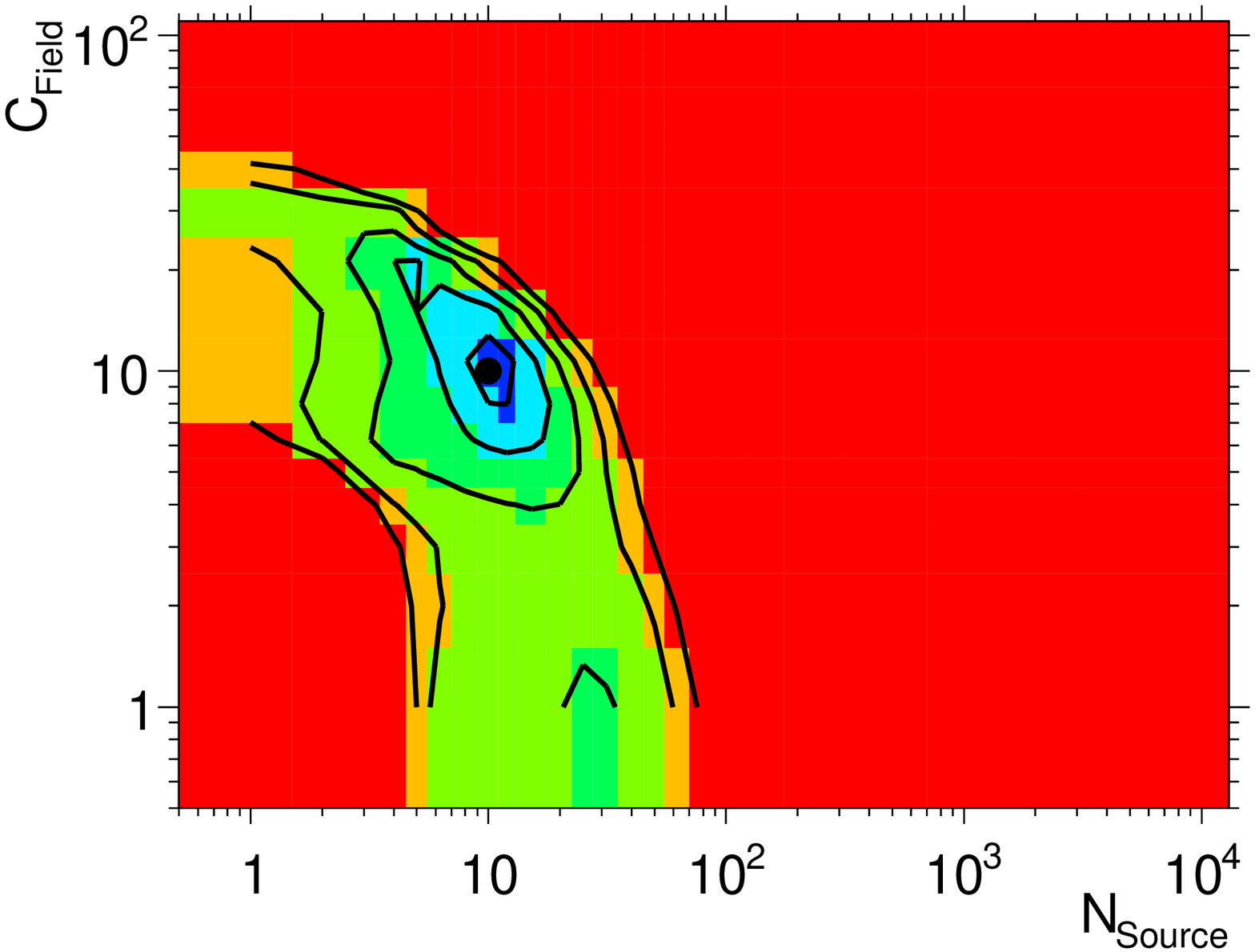}
 \label{fig:f3d}
 }
 
 \caption{\subref{fig:f3a} Energy-energy-correlations of the signal data (black square symbols) compared with the result of isotropic arrival directions (red triangle symbols), 
 \subref{fig:f3b} $\Omega$-distribution of the first bin of \subref{fig:f3a} (same color code),
 \subref{fig:f3c} Error contours for the reconstructed parameters resulting from the negative log-likelihood analysis (dark~blue~=~$1\sigma$-contour, light~blue~=~$2\sigma$-contour, dark~green~=~$3\sigma$-contour, light~green~=~$4\sigma$-contour, yellow~=~$5\sigma$-contour, red~$>$~$5\sigma$), the black point symbol shows the parameters of the signal data set,
 \subref{fig:f3d} Error contours using 100 signal data sets (same color code as in \subref{fig:f3c})}
 
\end{figure}

\subsection{Isotropic UHECR Arrival Directions} \label{sec:likelihood}

We compare the energy-energy-correlations of the signal data with the hypothesis of isotropic arrival. We simulated 100 UHECR data sets with isotropic arrival directions. For each set we calculated the energy-energy-correlations as described in section \ref{sec:analysis}. The average value of the resulting $\Omega$-distributions is shown in figure~\ref{fig:f3a} as the triangle symbols. The error bars represent the spread of the average values of the individual distributions. The square symbols show the same signal data distribution as in figure~\ref{fig:f2d}.

In order to quantify the agreement of the isotropic arrival model with the signal data distribution we use the negative log-likelihood method. For the comparison of the distributions we take bin-by-bin correlations into account. We make use of the approximately Gaussian shape of the $\Omega$ distributions within the angular $\alpha$ bins (e.g. figure~\ref{fig:f3b}), and define for the negative log-likelihood value
\begin{equation}\label{eq:ll}
L=-2 \ln \left( \frac{det(V^{-1})^{1/2}}{(2 \pi)^{n/2}} exp\left(-\frac{1}{2} (\boldsymbol{x}-\langle \boldsymbol{y} \rangle)^T \cdot V^{-1} \cdot (\boldsymbol{x}-\langle \boldsymbol{y} \rangle) \right)\right) .
\end{equation}

Here $\boldsymbol{x}$ and $\langle \boldsymbol{y} \rangle$ are n-dimensional vectors (n=10, the number of bins in figure~\ref{fig:f3a}) containing the signal data values of each $\alpha$ bin ($\boldsymbol{x}$), and the mean of the isotropic distributions ($\langle \boldsymbol{y} \rangle$). $V$ is the covariance matrix of the histogram resulting from the isotropic model (figure~\ref{fig:f3a}). The coefficients of $V$ of the distribution $\boldsymbol{y}$ are defined as follows,
\begin{equation}
 V_{ij}= \langle y_i y_j \rangle-\langle y_i \rangle \langle y_j \rangle.
\end{equation}

The resulting value of $L=295$ needs to be compared with an isotropic reference distribution in $L$.
Using isotropic distributions instead of the signal data we obtain a narrow distribution centered at $L_\circ=-46$ with a root mean square of $L_{RMS}=5$. These Numbers suggest that an isotropic UHECR scenario is excluded. In the analysis below we explicitly show that isotropy is excluded by more than five standard deviations (figure~\ref{fig:f3c}).

\subsection{Magnetic Field Model} \label{sec:scan}

In this section we confront a large number of variants of the magnetic model with the signal data distribution (figure~\ref{fig:f2d}). We vary the turbulent magnetic field as represented by the parameter $C_{Random Field}$ of equation \ref{eq:prop2}, and the number of sources $N_{source}$ emitting UHECRs with the same luminosity. The parameter for the regular magnetic field is kept fixed, as we found that the presented method has only minor sensitivity to variations of $C_{Coherent Field}$ at a fixed value of $N_{source}$. We kept the total number of UHECRs $N_{UHECR}=10000$ constant such that each source emits $N_{UHECR}/N_{source}$ UHECRs.

In order to cover the relevant regions of the parameter space we chose an almost logarithmic binning with 25 different numbers of sources (between 1 and 10000), and 16 different values of $C_{Random Field}$ (between 1~rad and 100~rad). For each of these values we simulated 100 data sets including the random walk and the coherent deflection, and determined the corresponding energy-energy-correlation distributions. The mean values $\langle \boldsymbol{y} \rangle$ and the covariance matrix $V$ are obtained for each pair of the model parameter values ($N_{source}, C_{Random Field}$).

For these values we calculated the negative log-likelihood values \linebreak $L(N_{source}, C_{Random Field})$ according to equation \ref{eq:ll} using for $\boldsymbol{x}$ again the signal data of figure~\ref{fig:f2d}. In order to reduce fluctuations resulting from the finite number of data sets in each ($N_{source}, C_{Random Field}$) bin, we smoothed the bins of the $L(N_{source}, C_{Random Field})$ distribution with a $5\times5$-kernel. The center weight of the kernel is $K_{33}=5$. All neighbor weights of $K_{33}$ have the weight $K_{22}=K_{23}=K_{24}=K_{32}=K_{34}=K_{42}=K_{43}=K_{44}=2$, and all second neighbors have zero weight except $K_{13}=K_{31}=K_{35}=K_{53}=1$.
This means the value of every bin is replaced by a weighted mean of the neighboring bins, according to the weights of the kernel.

The result of this approximation is shown in figure~\ref{fig:f3c}. The red area indicates the region of the model parameters which are excluded at the level of more than five standard deviations.
The evaluation of this model using the signal data distribution together with the log-likelihood method
therefore results in clear constraints on the allowed phase space of the model parameters.
The situation of an isotropic sky can be found at $N_{source}=10000$ sources
which is well within the excluded phase space region.

Beyond this, we have studied the reconstruction quality of the most likely parameter settings 
of the model under evaluation.
As the reference, the true input parameters of the signal data set are shown by the point symbol
in figure~\ref{fig:f3c}.

The contours in figure~\ref{fig:f3c} represent levels of $n$ standard deviations where
the inner dark blue region gives the result for $n=1$ standard deviation. 
The input parameter values are reconstructed within a region of two standard deviations. 
The shape of the distribution exhibits a continuous minimum region within the three standard 
deviation level.

Above the three standard deviation level, the contour features the following reduced sensitivity.
For the case of a small number of sources $N_{source}\le 2$ an enhanced 
strength of the random field component is able to partly simulate effects originating
from the $N_{source}=10$ source scenario.
The region of small turbulent fields cannot be excluded for the region around the 
original number of sources $N_{source}=10$.
Here the coherent component of the field with $C_{Coherent Field}=10~\mathrm{rad}$
gives a dominating effect.

For evaluating a given model using a measured data distribution of energy-energy-correlations,
the distribution shown in figure~\ref{fig:f3c} and its contours will be the main result.


For further crosschecks of the presented method, we produced 100 additional data sets with the same input parameters as the signal data set but with different random seeds. For each of these data sets the reconstruction of the parameters is repeated as described above.

In order to test the validity of the error contours we compared the reconstructed parameters with the input parameters of the model. We reconstructed the input parameter values for 40 data sets within one standard deviation, for 81 data sets within $2\sigma$, and for all 100 sets within $3\sigma$. These values are consistent with the expectation for a simultaneous determination of two parameters following a Gaussian distribution (39.3\% within $1\sigma$, 86.5\% within $2\sigma$, and 98.9\% within $3\sigma$).

We then averaged over the 100 uncertainty contours where the resulting distribution is shown in figure~\ref{fig:f3d}.
To investigate the potential of a systematic shift of the reconstructed parameters introduced by our method we checked the minimum of figure~\ref{fig:f3d} which lies within 1 standard deviation of the input parameters. This demonstrates that a possible systematic shift is small compared to the statistical precision of our procedure.

In addition we controlled the potential bias introduced by the specific choice of the random numbers we used to produce the initial signal data set. A simple example for such a bias would be two sources in the same line of sight, which would appear as one. Both figures~\ref{fig:f3c}, and ~\ref{fig:f3d} exhibit similar shapes of the allowed parameter space and demonstrate that our choice of the initial data set represents a typical realization of a UHECR sky with $N_{source}=10$ sources, and field strength of $C_{Random Field}=C_{Coherent Field}=10~\mathrm{rad}$.

\section{Conclusions}

In this paper we introduced an energy-energy-correlation observable $\Omega$ designed to obtain information on sources and propagation of UHECRs. 
The $\Omega$-distribution can be obtained directly from measured data without assumptions
on the underlying UHECR sky.

As models of UHECR emission and propagation become available they can be confronted with
the $\Omega$ measurements.
For quantitative evaluation of the models, we have presented a likelihood method.

To check the sensitivity of our procedure, we compared simple models with a simulated 
$\Omega$ measurement. 
We demonstrated how the phase space of model parameters, here number of sources and 
a parameter reflecting the average random magnetic field strength, 
were constrained using a large number of simulations where
these parameters were varied.


\section{Acknowledgments}
We wish to thank Johannes Erdmann for fruitful discussions, and a crosscheck of the random generators.
We are grateful to the Pierre Auger magnetic fields group and to the local group from Aachen for valuable discussions. We also thank Christopher Wiebusch for valuable comments on this publication. This work is supported by the Ministerium f\"ur Wissenschaft und Forschung, Nordrhein-Westfalen, and the Bundesministerium f\"ur Bildung und Forschung (BMBF).

\end{document}